\documentclass[twocolumn]{aastex62}
\usepackage{soul}
\usepackage{amsmath,amsfonts,amssymb}
\def\ADD#1{{\textcolor{black}{#1}}}
\usepackage{xcolor}
\def\be{\begin{equation}}
\def\ee{\end{equation}}
\def\ba{\begin{eqnarray}}
\def\ea{\end{eqnarray}}

\def\kp{k_\perp}

\def \pmbtext#1{\leavevmode
     \setbox0\hbox{#1}
     \kern0,4pt \copy0 \kern-\wd0
     \kern-0,2pt \raise0,3pt \box0 }

\shorttitle{$-8/3$ spectrum in the solar wind}
\shortauthors{David and Galtier}

\begin{document}
\title{$\kp^{-8/3}$ spectrum in kinetic Alfv\'en wave turbulence: implications for the solar wind}

\correspondingauthor{S\'ebastien Galtier}
\email{sebastien.galtier@u-psud.fr}

\author{Vincent David}
\affiliation{Laboratoire de Physique des Plasmas, \'Ecole polytechnique, F-91128 Palaiseau Cedex, France}
\affiliation{Univ. Paris-Sud, Observatoire de Paris, Univ. Paris-Saclay, CNRS, Sorbonne Univ.}

\author{S\'ebastien Galtier}
\affiliation{Laboratoire de Physique des Plasmas, \'Ecole polytechnique, F-91128 Palaiseau Cedex, France}
\affiliation{Univ. Paris-Sud, Observatoire de Paris, Univ. Paris-Saclay, CNRS, Sorbonne Univ.}
\affiliation{Institut universitaire de France}

\begin{abstract}   
The nature of solar wind turbulence at large scale is rather well understood in the theoretical framework of magnetohydrodynamics. 
The situation is quite different at sub-proton scales where the magnetic energy spectrum measured by different spacecrafts does not fit 
with the classical turbulence predictions: a power law index close to $-8/3$ is generally reported which is far from the predictions of strong and 
wave turbulence, $-7/3$ and $-5/2$ respectively. This discrepancy is considered as a major problem for solar wind turbulence. Here, we show 
with a nonlinear diffusion model of weak kinetic Alfv\'en wave turbulence where the cascade is driven by local triadic interactions \citep{Passot19}, 
that a magnetic spectrum with a power law index of $-8/3$ can emerge. This scaling corresponds to a self-similar solution of the second 
kind with a front propagation following the law $k_f \sim (t_*-t)^{-3/4}$, with $t<t_*$. This solution appears when we relax the implicit assumption 
of stationarity generally made in turbulence. The agreement between the theory and observations can be interpreted as an evidence of the 
non-stationarity of solar wind turbulence at sub-proton scales.  
\end{abstract}

\section{Introduction} \label{Intro}

The solar wind is a collisionless plasma characterized by fluctuations of its primary fields over a huge range of frequencies. One of the most 
spectacular properties reported from in situ measurements is a spectrum of magnetic fluctuations from frequencies $f \sim 10^{-6}$Hz to $\sim 100$Hz 
\citep{kiyani2015} with a spectral break around $f_b \sim 1$Hz \citep{Behannon78,Denskat83,Leamon98,Bourouaine12,Chen14}. 
This break separates the magnetohydrodynamic (MHD) scales ($f<f_{b}$) from the sub-proton scales ($f > f_{b}$) where ions and electrons are decoupled, 
and where signatures of kinetic Alfv\'en waves (KAW) can be found (see e.g. \cite{Sahraoui10,Salem12,Chen13}). 
\ADD{Note that signatures of other types of waves are also found (see e.g. \cite{Narita11,Roberts15}).}
Despite several years of studies, the nature of solar wind turbulence at sub-proton scales remains under debate 
(in this paper we restrict our attention to scales greater than the electron gyroscale). 
A reason is that the magnetic energy spectrum reported is generally close to $f^{-8/3}$ \citep{alexandrova12,Podesta13,Sahraoui13} which is far from 
the classical predictions of strong and (weak) wave turbulence 
\citep{biskamp96,GB03,galtier06,Galtier06b,Schekochihin09,Voitenko11,GM15,Cerri16,Passot18} for which the power law indices 
are $-7/3$ and $-5/2$ respectively. 

A debate is also developed around the mechanisms of energy dissipation. Although it seems necessary to heat the interplanetary 
collisionless plasma to explain its non-adiabatic cooling \citep{Richardson95}, the precise mechanism which involves kinetic effects is still not totally 
understood. For example we do not know if some dissipation occurs in the inertial range where \ADD{a spectrum close to $f^{-8/3}$ is found}. A possibility is that 
the latter power law is a spectrum predicted by a classical turbulence theory modified by some kinetic dissipation (see e.g. \cite{Passot15}). 
Note that several studies have been devoted to the question of solar wind heating and the evaluation of the energy cascade rate at MHD scales, 
which can be seen as a proxy for measuring the heating rate (see e.g. \cite{Sorriso2007,Vasquez2007,MacBride08,Osman11,Banerjee16,Hadid17}). 

The Letter is organized as follows. In section \ref{sec2} we introduce a model of KAW turbulence, first derived by \cite{Passot19}, and its phenomenology. 
In section \ref{sec3} we present its non-stationary solution which is a self-similar solution of the second kind. The numerical validation of our theory 
is given in section \ref{sec4}. A discussion is given in section \ref{sec5} about the applications to solar wind turbulence at sub-proton scales. 
A conclusion is finally proposed in the last section.

\section{Model of KAW turbulence}\label{sec2}

Nonlinear diffusion models are often used in the analysis of both strong \citep{Leith67,CN04,Matthaeus09,Thalabard15} and weak wave turbulence 
\citep{Zakharov99,Boffetta09,GNBT19}. There are mostly built by using phenomenological arguments but a rigorous treatment is 
sometimes possible in the regime of wave turbulence. The known examples are nonlinear optics \citep{Dyachenko} and MHD \citep{GB10}. In this 
case, the nonlinear diffusion equations are derived by taking the strongly local interactions limit of the kinetic equations; the latter equations being 
themselves derived in a systematical way. 
Recently, such a model has been proposed by \cite{Passot19} for KAW turbulence 
\ADD{(a model also valid for oblique whistler waves as explained in \cite{GM15})}
neglecting the coupling to other types of waves. 
The derivation can be qualified as semi-analytical because the problem is fundamentally anisotropic and in the final step of the derivation the authors 
neglected the cascade along the uniform magnetic field to find an expression for the nonlinear diffusion equation. However, the parallel cascade is 
expected to be relatively weak and its absence cannot be seen as a drawback of the model. Then, KAW turbulence is simulated numerically in 
presence of magnetic helicity in order to study the regime of imbalanced weak turbulence \citep{Passot19}. 
This type of model gives in general good quantitative information about the primary system because local interactions are in general the main driver 
of the turbulence cascade. 

Here, we shall use the diffusion equation proposed by \cite{Passot19} for weak KAW turbulence in absence of magnetic helicity. 
To be self-consistent (and for pedagogical reasons) a new derivation is proposed by using only phenomenological arguments. This method has the 
advantage of explaining in a simple way the main physical ingredients require to derive a nonlinear diffusion model for KAW turbulence. 

Since the leading nonlinear interaction of KAW is three-wave interaction \citep{GM15,Passot19}, the model is a second-order diffusion equation of the 
type
\be
\frac{\partial E(\kp)}{\partial t} = \frac{\partial}{\partial \kp} \left[ D_{\kp} E(\kp) \frac{\partial (E(\kp) / \kp)}{\partial \kp} \right] \, ,
\label{DE1}
\ee
where $E(\kp)$ is a one-dimensional magnetic energy spectrum, $\kp$ the perpendicular wavenumber and $D_{\kp}$ a diffusion coefficient that eventually 
depends on the wavenumber $\kp$. This equation is constructed in such a way that it preserves the nonlinearity degree with respect to the spectrum 
(quadratic in our case) and, its cascade and thermodynamic solutions. We neglect the cascade along the strong uniform magnetic field ${\bf B_{0}}$
which defines the parallel direction, hence the presence of only the perpendicular (to ${\bf B_{0}}$) wavenumber $\kp$. A dimensional analysis of 
expression (\ref{DE1}) gives
\be
\frac{E(\kp)}{\tau} \sim \frac{ D_{\kp} E^2(\kp)}{\kp^3} \, , 
\ee
where $\tau$ is the cascade time of weak wave turbulence; thus 
\be
D_{\kp} \sim \frac{\kp^3}{\tau E(\kp)} \sim \frac{\kp^3}{(\tau_{KAW}/\epsilon^{2}) E(\kp)} \, .
\ee
The KAW time is given by the relation 
\be
\tau_{KAW} \sim \frac{1}{\omega} \sim \frac{1}{k_\parallel \kp} \sim \frac{1}{\kp}  \, ,
\ee
where $\epsilon \sim \tau_{KAW} / \tau_{NL} \ll 1$ is a small parameter and the nonlinear time $\tau_{NL} \sim 1 / (\kp^2 \sqrt{\kp E(\kp)})$.
We obtain
\be 
D_{\kp} \sim \kp^7 \, ,
\label{diff1}
\ee
which leads to the following second-order diffusion equation for KAW turbulence \citep{Passot19} 
\be
\frac{\partial E(\kp)}{\partial t} = C \frac{\partial}{\partial \kp} \left[ \kp^7 E(\kp) \frac{\partial ( E(\kp)/\kp )}{\partial \kp} \right] \, ,
\label{DE0}
\ee
where $C$ is a positive constant. 

The constant flux solutions can now be found. We define the energy flux $\Phi_E(\kp)$ as follows
\be
\frac{\partial E(\kp)}{\partial t} = - \frac{\partial \Phi_E (\kp)}{\partial \kp} \, 
\ee
and introduce the magnetic energy spectrum $E(\kp) = A \kp^x$ into equation (\ref{DE0}) with $A$ a positive constant. We find
\be
\Phi_E(\kp) = A^2 C (1-x) \kp^{5+2x} \, . 
\label{fluxE}
\ee
The constant flux solutions are $x=1$, which corresponds to the thermodynamic solution (zero flux), and $x=-5/2$ called the Kolmogorov-Zakharov solution
\citep{nazarenko11}. In this case we also find $\Phi_E(\kp) \equiv \Phi_0 = (7/2) A^2 C$ which is positive and thus corresponds to a direct cascade. 
Therefore, we recover the well-known solutions of the problem \citep{GB03,GM15,Passot19}.

\section{Non-stationary regime}\label{sec3}

Time-dependent solutions of the KAW turbulence equation (\ref{DE0}) will be studied further analytically and numerically. We will demonstrate the 
existence of a non-trivial solution \ADD{(called sometimes anomalous scaling)} in the sense that it cannot be derived with the usual turbulence 
phenomenology or theory. This property is related to the finite capacity of the system which is linked to the convergence of the integral
\be
\int_{k_i}^{+\infty} E(\kp) d\kp \, , 
\ee
where $k_i$ is the scale of magnetic energy injection. This property is satisfied when $x<-1$, a situation found in KAW turbulence. 

The non-stationary spectrum can be modeled as a self-similar solution of the second kind (see e.g. \cite{Falkovich91,Thalabard15}) 
taking the form
\be
E(\kp) = \frac{1}{\tau^{a}} E_0 \left(\frac{\kp}{\tau^b} \right) \, , 
\ee
where $\tau = t_*-t$, and $t_*$ is a finite time at which the magnetic energy spectrum reached the largest available wavenumber. 
By introducing the above expression into (\ref{DE0}) we find the condition
\be
a=4b+1 \, . 
\ee
A second condition can be found by assuming that $E_0(\xi) \sim \xi^m$ far behind the front. Then, the stationarity condition gives the following relation
\be
a + mb = 0 \, . 
\ee
Finally, the combination of both relations gives 
\be
m=-\frac{a}{b} = -4 - \frac{1}{b} \, . 
\ee
The latter expression means that we have a direct relation between the power law index $m$ of the spectrum and the law of the front propagation 
which follows $k_f \sim \tau^b$. For example, if we assume that the stationary solution -- the Kolmogorov-Zakharov spectrum -- is established 
immediately during the front propagation, then $m=-5/2$ and $b = -2/3$ (and $a=-5/3$). In this case, the prediction for the front propagation is 
\be
k_f \sim (t_*-t)^{-2/3} \, . 
\ee

\section{Numerical simulation}\label{sec4}
We now study numerically the time evolution of the magnetic energy spectrum described by the KAW turbulence equation (\ref{DE0}) with $C=1$. 
A linear hyper-viscous term of the form $-\eta \kp^6 E(\kp)$ is added to equation (\ref{DE0}) in order to introduce a sink at small scale to avoid the 
development of numerical instabilities at the final time of the simulation ($t>t_*$) when the stationary state establishes; we take $\eta = 10^{-16}$. 
A logarithmic subdivision of the $\kp$-axis is used with ${\kp}_i= 2^{i/8}$ and $i$ an integer varying between $0$ and $160$. Note that this resolution 
is far too large to model the sub-proton scales where electron inertia is neglected. This choice is however necessary to reach a clear conclusion about 
the values of the power law indices (see below). The Crank-Nicholson and Adams-Bashforth numerical schemes are implemented for the nonlinear and 
dissipative terms respectively. The initial condition ($t=0$) corresponds to a spectrum localized at large scale with $E(\kp) \sim \kp^3 \exp(-(\kp / k_0)^2)$ 
and $k_0=5$. No forcing is added at $t>0$. The time-step is $dt=2 \times 10^{-13}$. 

\begin{figure}
\begin{centering}
\includegraphics[width=1\linewidth]{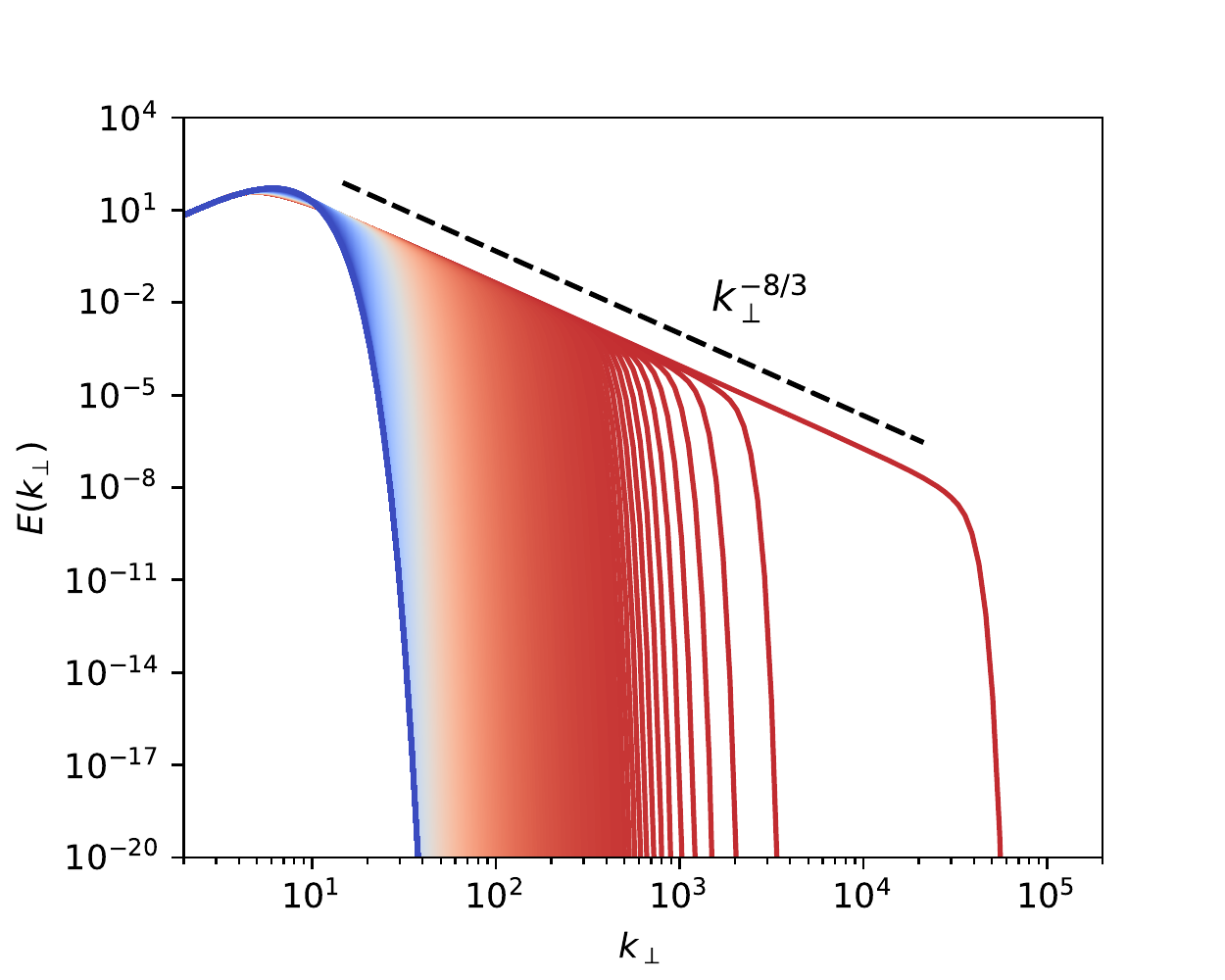}\par
\end{centering}
\caption{Time evolution (every $1000 dt$) of the magnetic energy spectrum $E(\kp)$ from $t=0$ (blue) to $t_*$ (dark red). 
A $\kp^{-8/3}$ spectrum emerges over three decades.}
\label{Fig1}
\end{figure}
\begin{figure}
\begin{centering}
\includegraphics[width=1\linewidth]{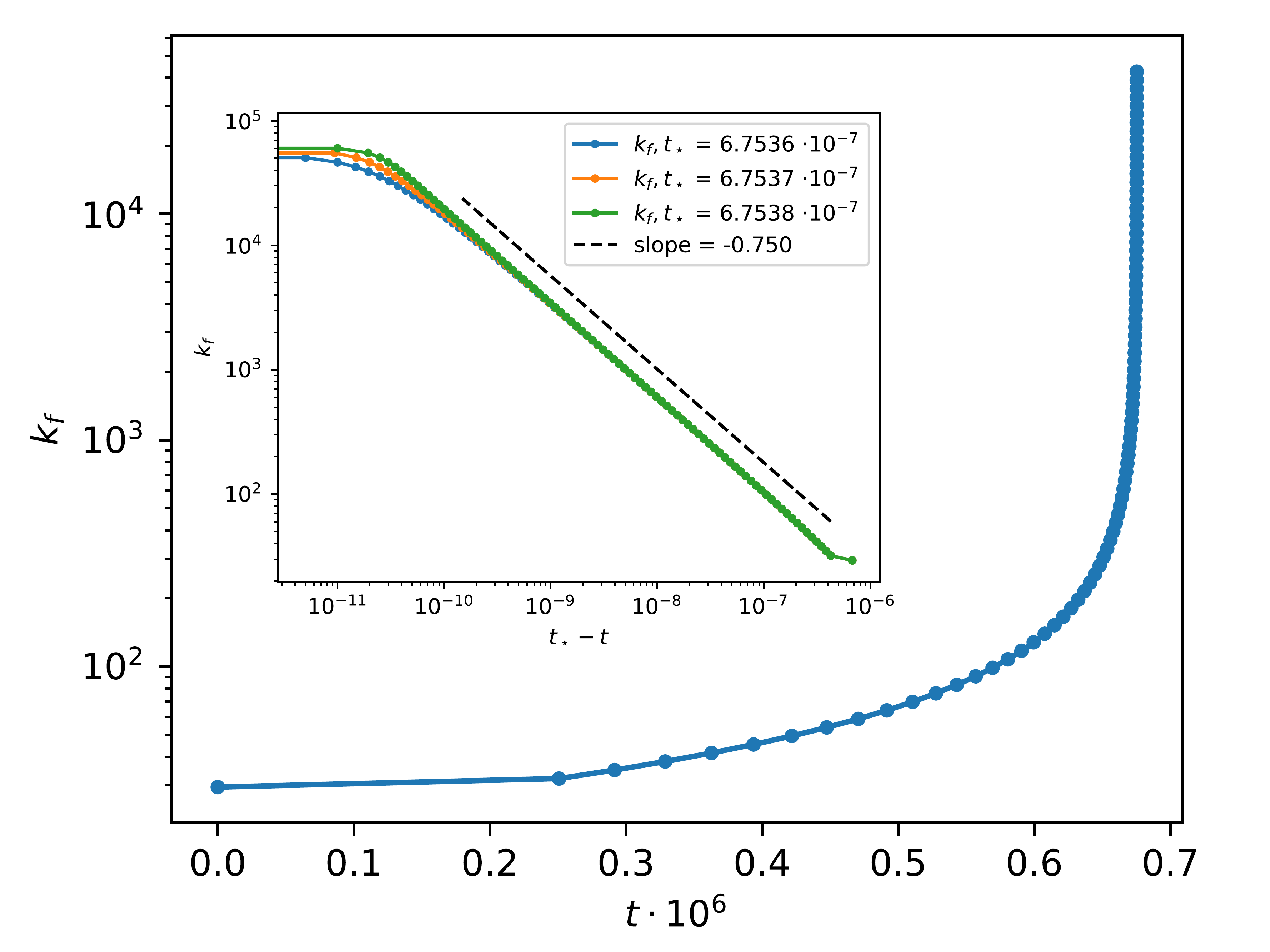}\par
\end{centering}
\caption{Temporal evolution of the spectral front $k_f$ for $t \le t_*$ in linear-logarithmic coordinates (blue). A sharp increase of $k_f$ is observed from 
which we can define precisely the singular time $t_* = 6.7537 \times 10^{-7}$. Inset: The temporal evolution of $k_f$ as a function of $t_*-t$ (orange) in 
double logarithmic coordinates. The black dashed line corresponds to $(t_*-t)^{-0.750}$. For comparison two other values of $t_*$ are taken (green and 
blue).}
\label{Fig2}
\end{figure}
\begin{figure}
\begin{centering}
\includegraphics[width=1\linewidth]{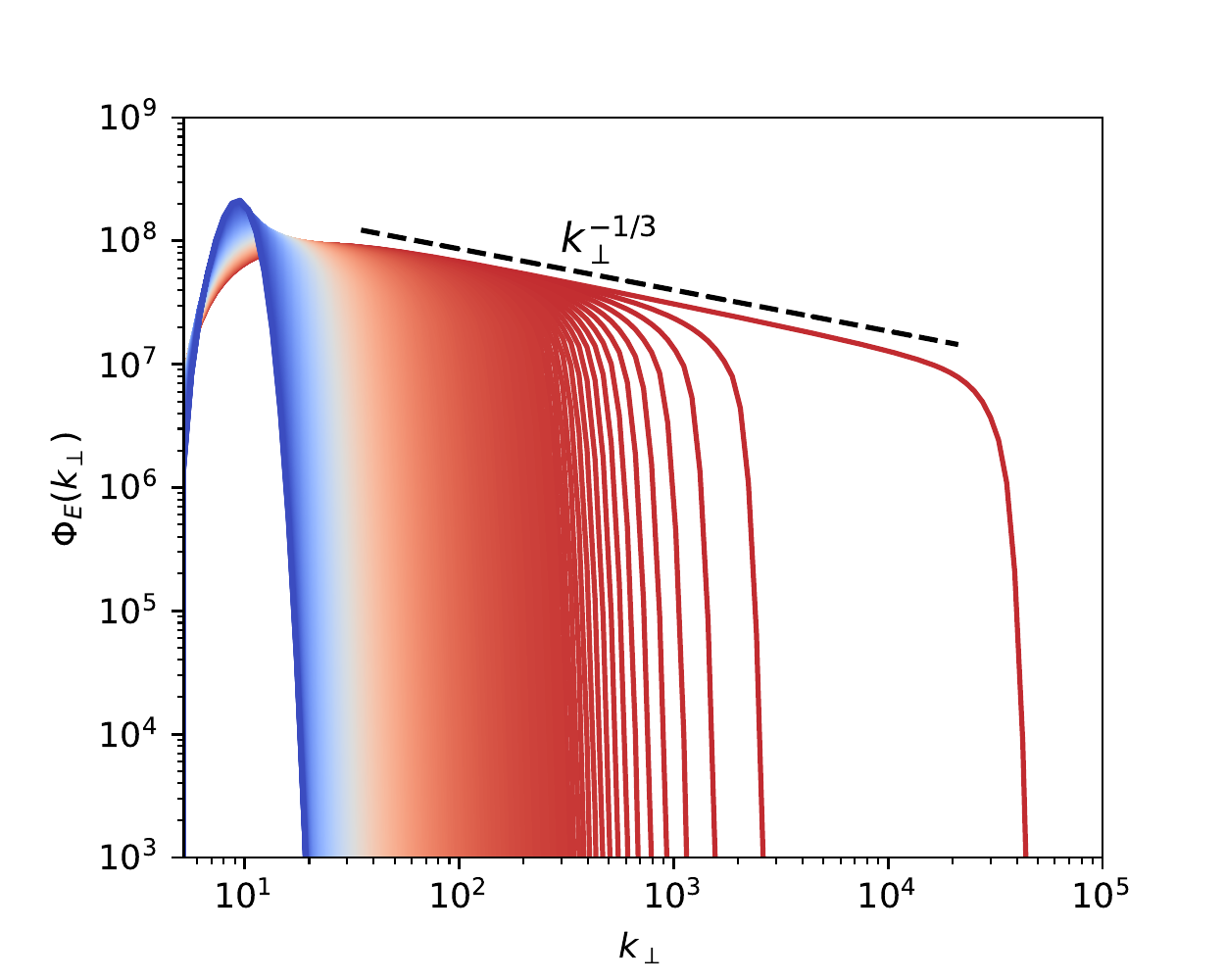}\par
\end{centering}
\caption{Temporal evolution of the energy flux $\Phi_E(\kp)$ in double logarithmic coordinates for the same times as in Figure \ref{Fig1} (same conventions).
The flux follows a power law $\sim \kp^{-1/3}$.}
\label{Fig3}
\end{figure}
\begin{figure}
\begin{centering}
\includegraphics[width=1.1\linewidth]{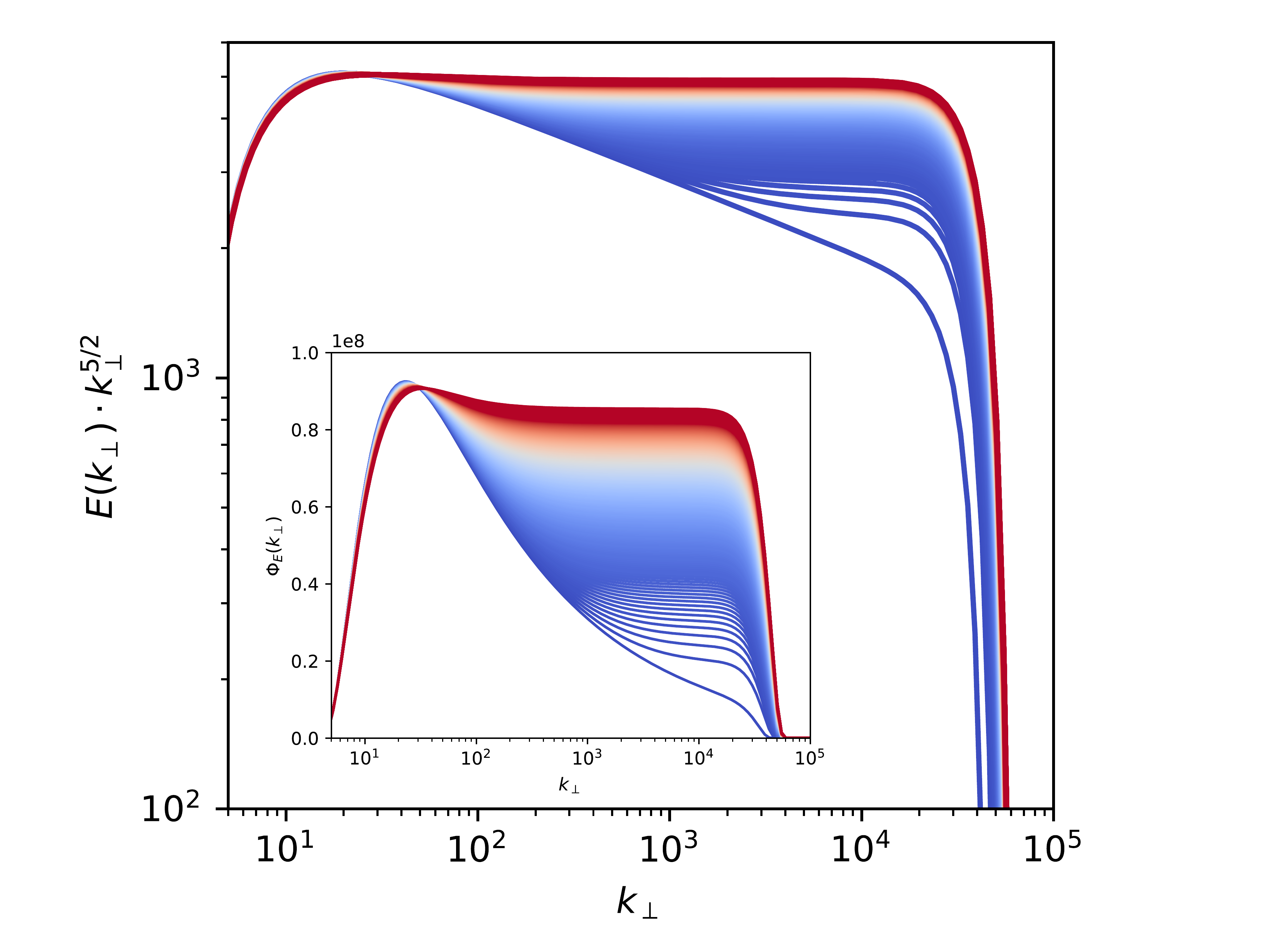}\par
\end{centering}
\caption{Temporal evolution (every $1000 dt$) for $t> t_*$ of the energy spectrum compensated by $\kp^{5/2}$ (in double logarithmic coordinates). 
Inset: Temporal evolution of the energy flux $\Phi_E(\kp)$ for the same times (in linear-logarithmic coordinates).}
\label{Fig4}
\end{figure}

In Figure \ref{Fig1} we show the time evolution of the magnetic energy spectrum from $t=0$ to $t_{*}$. During this non-stationary phase a clear power law
spectrum in $\kp^{-8/3}$ is formed behind the front. To check if this spectrum corresponds to the self-similar solution of the second kind introduced above
we show in Figure \ref{Fig2} the front propagation $k_{f}(t)$. This front is defined by taking $E(\kp)=10^{-15}$ from Figure \ref{Fig1}: we then follow the point of 
intersection between this threshold and the spectral tail. From Figure \ref{Fig2} we can define the singular time $t_*$ at which the front can reach in 
principle $\kp = + \infty$. Note that a similar situation where the small scales are reached in a finite time is also observed e.g. in Alfv\'en wave turbulence 
\citep{galtier00}. The value $t_* = 6.7537 \times 10^{-7}$ is chosen. 
In Figure \ref{Fig2} (inset) we show $k_f$ as a function of $t_*-t$: a clear power law is observed over three decades with a power law index of $-0.750$. 
The negative value illustrates the explosive character of the direct cascade of magnetic energy in KAW turbulence. 
The different values measured are fully compatible with 
\be
a=-2 \, , \quad b=-3/4 \quad \rm{and} \quad m=-8/3 \, ,
\ee
which therefore demonstrates the self-similar nature of the non-stationary solution. 

As displayed in Figure \ref{Fig3}, the non-stationary phase is characterized by a non-constant energy flux $\Phi_E(\kp)$ (computed from the nonlinear terms): 
we start with a flux localized 
at small wavenumbers which then develops towards smaller scales without reaching a plateau. The solution does not correspond to the constant 
flux solution derived analytically, but it is fully compatible with the power law solution $\sim \kp^{-1/3}$ when we take $x=-8/3$ in Equation (\ref{fluxE}). 

Finally, in Figure \ref{Fig4} we show the temporal evolution for $t>t_*$ of the energy spectrum and energy flux (inset), respectively. The classical 
(stationary) wave turbulence predictions are finally obtained with an energy spectrum in $\kp^{-5/2}$ and a constant positive energy flux, as expected 
for a direct cascade. This behavior is specific to a viscous simulation made in a finite box where the cascade cannot continue to smaller scales: 
the energy accumulates at small scale until the viscous term (proportional to the energy spectrum) becomes non-negligible and balance the 
energy flux coming from large scale. This process affects the entire inertial range with a modification of the power law index. 
The final phase of the simulation (not shown) corresponds to a self-similar decay of the energy spectrum with the same power law index ($-5/2$).

\section{Solar wind turbulence at sub-proton scales}\label{sec5}
Solar wind turbulence at sub-proton scales \ADD{(for frequencies $f>1$Hz)} is characterized by a magnetic energy spectrum with a power-law index 
close to $-8/3$ \citep{alexandrova12,Podesta13,Sahraoui13}. This scaling law does not correspond to the classical prediction of strong turbulence 
($-7/3$) or weak wave turbulence ($-5/2$), which are obtained phenomenologically or analytically respectively, with different types of model equations, 
and with different types of waves, in presence of anisotropy or not. 
After several years of investigations, the possibility of having a power law index close to the data seemed to be impossible with the classical 
turbulence theory (see however \cite{Boldyrev12,Meyrand13}). For this reason, this problem is one of the most important in space plasma physics. 
A natural conclusion is that the observed power laws are the result of a non-trivial turbulent dynamics that we still do not understand or a physics 
involving ingredients other than turbulence. 

In this Letter, we have shown with a nonlinear diffusion model of weak KAW turbulence, which retains only local interactions \citep{Passot19}, 
that by relaxing the implicit assumption of stationarity generally made in turbulence to obtain predictions, a new solution -- a self-similar solution 
of the second kind -- is possible for KAW turbulence. It is characterized by a magnetic energy spectrum in $\kp^{-8/3}$ which coincides with 
in situ observations. In this non-stationary phase the viscous dissipation is negligible. 
While the absence of viscous dissipation should be considered as the right way to tackle the problem of solar wind turbulence at sub-proton scales, 
since the solar wind is a collisionless plasma and thus cannot behave like a viscous fluid, we must nevertheless clarify the meaning and the 
consequences of such assumption. The first clear idea is that there is no reason to believe that dissipation at kinetic scales should behave like that 
found in hydrodynamics; Landau damping is a good example. 
According to our interpretation the results obtained here are in favor of a kinetic dissipation that does not produce a feedback 
on the inertial range of KAW turbulence. This property is at odds with fluid turbulence. 
We might also conclude that the kinetic dissipation is simply negligible, however, the presence of kinetic dissipation as a source of 
plasma heating seems to be necessary to explain the slow (ion) temperature variation with the heliocentric distance \citep{Richardson95}. 
According to our study, we can also think that the observation of \ADD{a spectral index close to $-8/3$} in the solar wind is a consequence of the 
existence of a cascade at electron scale since in this case the accumulation of magnetic energy found in our simulation is not favored. 
The physics at electron scales is, however, quite different: for example the magnetic energy is not an invariant anymore (see e.g. \cite{Meyrand10}). 
Then, the feedback of these scales on the ion scales studied in this paper is non-trivial. 
\ADD{Note finally that the weak turbulence regime studied in this Letter also provides a natural explanation to the enigmatic non-Gaussian 
mono-scaling observed at sub-proton scales \citep{kiyani09}.}

\section{Conclusion}\label{sec6}
Our study reveals that the classical hypothesis of stationarity to obtain any turbulence predictions may not be the best way to understand solar wind 
turbulence at sub-proton scales. Instead, the relaxation of this assumption opens a new type of solution that is understood as a self-similar solution of 
the second kind. On the basis of numerical simulations of a nonlinear diffusion model of weak KAW turbulence we show that the main scaling behavior 
observed with spacecrafts -- a power law index close to $-8/3$ for the magnetic energy spectrum -- which has so far resisted classical theoretical modeling, 
can be reproduced with a fairly high accuracy. 
%
%
The non-stationary nature of solar wind turbulence at sub-proton scales can be explained by an imbalance between nonlinearities and kinetic 
dissipation, and by the existence of a cascade at electron scales. 
The nature of the kinetic dissipation in a collisionless plasma remains to be explained in details to fit these constraints. 

The solar wind is the best example for studying unbounded collisionless plasmas in astrophysics. It is quite challenging to understand the 
behavior of such plasmas in the regime of turbulence but surprisingly some simplicity seems to emerge from complexity with a fluid-like behavior 
(see e.g. \cite{Meyrand19,Wu19}). Space missions like Parker Solar Probe and Solar Orbiter could also help in testing theoretical ideas.

\end{document}